\documentclass[12pt,english,dvips]{scrartcl}
\usepackage{babel}
\usepackage[T1]{fontenc}
\usepackage[ansinew]{inputenc}

\begin{document}

\begin{titlepage} \vspace{0.2in} 

\begin{center} {\LARGE \bf 

Geometrization of the Gauge Connection within a
Kaluza-Klein Theory}\\ 
\vspace*{1cm}
{\bf Giovanni Montani}\\
\vspace*{1cm}
ICRA---International Center for Relativistic Astrophysics\\ 
Dipartimento di Fisica (G9),\\ 
Universit\`a  di Roma, ``La Sapienza",\\ 
Piazzale Aldo Moro 5, 00185 Rome, Italy.\\ 
e-mail: montani@icra.it\\ 
\vspace*{1.8cm}

 \vspace*{1cm} \\

{\bf   Abstract  \\ } \end{center} \indent
Within the framework of a Kaluza-Klein theory,
we provide the geometrization
of a generic (Abelian and 
non-Abelian) gauge coupling, which comes out by
choosing a suitable matter fields dependence
on the extra-coordinates.\\

We start by the extension of the N\'other
theorem to a multidimensional spacetime being the
direct sum of a 4-dimensional Minkowski space and
of a compact homogeneous manifold (whose isometries reflect
the gauge symmetry); we show, how on such a
``vacuum'' configuration, the extra-dimensional
components of the field momentum correspond to
the gauge charges.
Then we analyze the structure
of a Dirac algebra as referred to a spacetime with the
Kaluza-Klein restrictions and,
by splitting
the corresponding free-field Lagrangian, we
show how the gauge coupling terms outcome. 

\end{titlepage}

\section{BASIC STATEMENTS}

The works of Kaluza \cite{K21} and Klein
\cite{K26,K26a}
allowed to include the electromagnetic field
within a geometrical picture, by adding an extra
spacelike dimension to the spacetime;
in spite of this success, the Kaluza-Klein theories
had their full development only after the
formulation of non-Abelian gauge theories
\cite{MS84,C88}. In fact the main achievement
of this approach
(for a complete discussion about this topic,
see the works collected in \cite{ACF87} or the
review presented in \cite{OW97})
relies on the
geometrization of Yang-Mills fields, whose group of
symmetry admits a representation in terms of
an isometry in the extra-dimensions
\cite{C175,C275}
(see also \cite{C78} for an extension of the
multidimensionality idea 
to the supergravity theory).\\
The price that the Kaluza-Klein theories have to play
for such a geometrical picture of unification,
consists of restrictions on the physically admissible
spacetime an coordinates transformations. 
In fact, the spacetime has to take the structure of a
generic 4-dimensional manifold plus a compact
homogeneous hypersurface (such a feature implies a
violation of the {\em Equivalence Principle}) and
along the extra-dimensions only translations of the
coordinates are available (so violating the
{\em Principle of General Relativity} as extended to
a multidimensional spacetime).\\
However the geometrical theories of unification
can be settled down within General Relativity by means of
the so-called {\em Spontaneous Compactification}
process \cite{W81}. In this framework the Lagrangian of the
theory is yet the multidimensional
Einstein-Hilbert one, but we observe the
Kaluza-Klein restriction because the ``vacuum state''
has the compactified structure
(i.e. it is a 4-dimensional Minkowski space
plus a compact homogeneous manifold); thus we may
interpret the dimensional compactification as a
spontaneous breaking of the Poincar\'e symmetry.\\
Aim of the present analysis consists of extending
the Kaluza-Klein approach even to the gauge connection
associated to the Yang-Mills fields; it is achieved
by splitting a free multidimensional spinor field
Lagrangian and fixing the hypotheses necessary
for the appearance
(within the reduced 4-dimensional action) 
of the gauge connection terms.
 
While the two Sections 2 and 3 are devoted to review
respectively the gauge theories and the Kaluza-Klein
approach, in Section 4,
as first step of this work,
we generalize the
N\"other theorem to the vacuum of a Kaluza-Klein
theory; we show that the extra-dimensional
components of the momentum operator correspond to the
conserved charges associated to the gauge symmetry.
Such an identification requires suitable hypotheses
on the ``matter'' fields dependence with respect to
the extra-coordinates; we take such a dependence in
the form of a phase factor which include the gauge
generators. In Section 5 we analyze the Dirac algebra
on a Kaluza-Klein spacetime and see how the
$\gamma$-matrices relations are preserved by the
dimensional reduction process. The tangent space
projection of the spinor connection is discussed
to outline that its extra-dimensional component is
a free quantity of our problem.

In Section 6,
on the basis of a Lagrangian approach,
we split the action of a free multidimensional spinor
field and, after the dimensional reduction
(here is crucial to take the integral over the
extra-coordinates), we get the action of a
4-dimensional free spinor field plus the gauge
coupling with the Yang-Mills fields;
undesired terms appearing in the splitted action are
removed by fixing the residual spinor connection
components.\\
In Section 7 brief concluding remarks are provided
to stress a difference existing between the Abelian
and non-Abelian case.

\section{GAUGE THEORIES}

Let us assign, on the Minkowski space ${\cal M}^4$
(endowed with the coordinates system
$\{ x^{\mu }\} \; \mu = 0,1,2,3$),
a set of fields
$\phi _r(x^{\mu })$ ($r=1,2,...,n\in N$), whose Lagrangian density
${\cal L}(\phi_r, \; \partial _{\mu }\phi _r)$ is invariant
under the unitary transformations 

\begin{equation}
\phi _r = \left( exp\{ ig\omega ^aT_a\} \right) _{rs}\phi _s
\, ,
\label{a}
\end{equation}

where $\omega ^a$ ($a=1,2,...,K\in N$) denote constant parameters.
while $T_{a\; rs}$ are the $K$-dimensional
(Hermitian) group generators
corresponding to the coupling constant $g$.\\ 
In agreement with the N\"other theorem \cite{MS84,C88}, this invariance implies
the existence of the conserved charges

\begin{equation}
Q_{a} = ig\int_{{\cal E}^3} d^3x\pi _rT_{a\; rs}\phi _s 
\, , 
\label{b}
\end{equation}

being $\pi _r$ the conjugate momentum to $\phi _r$ and
${\cal E}^3$
the 3-dimensional Euclidean space.
We upgrade the global symmetries (\ref{a}), to a
(local) gauge one, by requiring that their parameters become
spacetime functions, i.e.
$\omega ^a = \omega ^a(x^{\mu })$. The invariance of the
theory (i.e. of the Lagrangian) under the
gauge symmetries involves
new fields $A^a_{\mu }(x^{\nu })$ into the dynamics
(the so-called Yang-Mills fields);
such fields, under infinitesimal gauge transformations
($\omega ^a$ is replaced by $\delta \omega ^a\ll 1$), behave as

\begin{equation}
A^a_{\mu }\rightarrow  A^a_{\mu } 
-\varepsilon ^{abc}\delta \omega ^bA^c_{\mu } -
\partial _{\mu }\delta \omega ^a
\, ; 
\label{c}
\end{equation}

here the quantities $\varepsilon ^{abc}$ denote the structure
constants of the Lie group and are defined via the relation
$[T_{a},\; T_{b}]=i\varepsilon ^c_{ab}T_{c}$
(indices $a,b,c$ are raised and lowered in Euclidean sense
and repeated ones are summed from $1$ to $K$).
The fields $A^a_{\mu }$ are said Abelian or non-Abelian whether
the structure constants vanish or not.\\
In gauge invariant form, the Lagrangian density of the theory
rewrites as 

\begin{equation}
{\cal L}(\phi _r,\; D_{\mu }\phi _r) - 
\frac{g^2}{4} F^{a\; \mu \nu }F_{a\; \mu \nu }
\, ;
\label{d}
\end{equation}

above
$D_{\mu }\phi _r\equiv \partial _{\mu }\phi _r
+ igT_{a\; rs}A^a_{\mu }\phi _s$
indicates the {\em gauge covariant derivative} and
the quadratic term in the gauge tensors 
$F^a_{\mu \nu} \equiv \partial _{\nu }A_{\mu } -
\partial _{\mu }A_{\nu }
+ \varepsilon ^{abc}A^b_{\mu }A^c_{\nu }$
provides the gauge vector fields dynamics.

\section{KALUZA-KLEIN PARADIGM}

Within a Kaluza-Klein theory \cite{ACF87}, the gauge fields are
geometrized by adding, to the 4-dimensional
spacetime
${\cal V}^4$ (having internal coordinates
$x^{\gamma }\; \gamma = 0,1,2,3$), 
a compact homogeneous D-dimensional space $\Sigma ^D$
(having a very  small size and adapted coordinates
$\theta ^l \; l = 4,...,D$, whose isometries corresponds to
the gauge symmetries.
If we take the dimension of the gauge group equal to that one
of the extra-space (i.e. $K=D$), then the whole manifold
${\cal V}^{4+D} = {\cal V}^44\times \Sigma ^D$ admits, in the
4+D-bein representation. the line element

\begin{equation}
ds^2 = \eta _{(C)(D)} e^{(C)}_A e^{(D)}_BdX^AdX^B \, \quad 
\eta _{(C)(D)} = diag \{1,-1,...,-1\} 
\, ,
\label{e}
\end{equation}

being $X^A \; , A,B, = 0,1,...,D$ the coordinates on
${\cal V}^{4+D}$ (i.e. $X^A = \{ x^{\mu },\; \theta ^l\} $), 
while the indices in parenthesis refer to the 4+D-bein;
here the vectors $e^{(C)}_A$ take the form

\begin{eqnarray}
\label{f} 
e^{(A)}_{\mu } = \left( u^{(\nu )}_{\mu }(x^{\gamma }) \; ; \; 
\gamma A^{(n)}_{\mu }(x^{\gamma })\right) \\
e^{(A)}_m = \left( {\bf 0} \; ; \; 
\xi ^{(n)}_m(\theta ^l)\right) 
\, . 
\end{eqnarray}

The reciprocal vectors $e_{(C)}^A$
(such that $e^{(C)}_Ae^B_{(C)}=\delta ^B_A\; ; \; 
e^{(C)}_Ae^A_{(D)}=\delta ^{(C)}_{(D)}$) are as follows

\begin{eqnarray}
\label{g} 
e^{A}_{(\mu )} = \left( u^{\nu }_{(\mu )}(x^{\gamma }) \; ; \; 
-\gamma A^{(n)}_{\nu }(x^{\gamma })u^{\nu }_{(\mu )}(x^{\gamma })
\xi^m_{(n)}(\theta ^l)\right) \\
e^{A}_{(m)} = \left( {\bf 0} \; ; \; 
\xi ^{n}_{(m)}(\theta ^l)\right) 
\, ; 
\end{eqnarray}

here the vectors $u^{\nu }_{(\mu )}$ and $\xi ^n_{(m)}$ are
reciprocal respectively to
$u^{(\nu )}_{\mu }$ and $\xi ^{(n)}_m$, while $\gamma = const.$. 

The vectors $\xi ^{(m)}_n$ correspond to the Killing fields
of the compact manifold $\Sigma ^D$ and therefore satisfy
the relations

\begin{eqnarray}
\label{h} 
\partial _n\xi^{(p)}_m -
\partial _m\xi^{(p)}_n =
C^{(p)}_{(q)(r)} \xi ^{(q)}_m\xi^{(r)}_n\\
~^D\nabla _n\xi^{(p)}_m +
~^D\nabla _m\xi^{(p)}_n = 0 
\, , 
\end{eqnarray}

where
$C^{(p)}_{(q)(r)}$ denote the structure constants of
the isometries group on $\Sigma ^D$ and 
the covariant derivative $~^D\nabla _m$
refers to the extra-dimensional metric.

The Kaluza-Klein paradigm is implemented
by requiring the explicit symmetry breaking of the
4+D-dimensional diffeomorphisms into the 4-dimensional
general coordinates transformations
$x^{\mu ^{\prime }} =
x^{\mu ^{\prime }}(x^{\nu })$ 
and a translation along the extra-coordinates
$\theta ^{m^{\prime }} = \theta ^m + 
\omega ^{(p)}(x^{\nu })\xi ^m_{(p)}$.
Under infinitesimal transformations of the above type, 
$u^{(\nu )}_{\mu }$
behave like 4-bein vectors, $\xi ^{(n)}_m$ like scalar
quantities and $A^{(m)}_{\mu }$ transform according to
gauge 4-fields.\\
The geometrization of a gauge group is achieved by requiring
that its structure constants $\varepsilon ^{abc}$
and the coupling constant $g$ 
coincide, respectively, with those ones 
of the isometries $C^{(p)(q)(r)}$
and with the constant
$\gamma \sqrt{c^3/16\pi ~^4G}$
(where the 4-dimensional Newton constant reads from the
multi-dimensional one as 
$~^4G\equiv ~^{4+D}G/V$, being $V$ the volume of $\Sigma ^D$).
In fact, hence the 4+D-dimensional
Einstein-Hilbert action provides, after dimensional
reduction,
the 4-dimensional Einstein-Yang-Mills one
(i.e.
we get ordinary 4-gravity and a Yang-Mills contribution).

\section{N\"OTHER THEOREM} 

We discuss the invariance of a 4+D-dimensional field theory,
on the spacetime
${\cal M}^4\times \Sigma ^D$, 
(in absence of gauge fields, i.e. $A^{(m)}_{\mu }\equiv 0$)  
under a translation of the coordinates, i.e. we extend
the N\"other theorem \cite{MS84}  to the
extra-dimensional context.\\
Let us consider a set of fields
$\varphi _r(x^{\mu }, \theta ^m)$ ($r = 1,2,...,n$), 
whose Lagrangian density ${\cal L}$ is
invariant under the infinitesimal coordinates displacement
$x^{\mu ^{\prime }} = x^{\mu } + \delta \omega ^{\mu }$
and
$\theta ^{m^{\prime }} = \theta ^m +
\delta \omega ^{(p)}\xi ^m_{(p)}$ with
$\left( \delta \omega ^{\mu }\;
\delta \omega ^{(p)}\right) = const.$;
in 4+D-dimensional notation, we take the infinitesimal
coordinates transformation
$X^{A^{\prime }}= X^A + \delta \omega ^{(B)}e^A_{(B)}$
(with $\delta \omega ^{(A)} =
\left( \delta \omega ^{\mu }, \; \delta \omega ^{(p)}\right) $
and $u^{\nu }_{(\mu )} = \delta ^{\nu }_{(\mu )}$) which,
in turn, induces the corresponding fields transformation

\begin{equation}
\varphi ^{\prime }_r = \varphi _r + \delta \varphi _r
\, \quad ,\; 
\delta \varphi _r = \partial _A\varphi _re^{A}_{(B)}
\delta \omega ^{(B)} 
\, . 
\label{i} 
\end{equation}

The Lagrangian density invariance provides

\begin{equation}
\partial _A{\cal L} e^{A}_{(B)}\delta \omega ^{(B)} =
\partial _A\left(
\frac{\partial {\cal L}}{\partial (\partial _A\varphi _r)} 
\delta \varphi _r\right) - 
\left[ \partial _A\left(
\frac{\partial {\cal L}}{\partial (\partial _A\varphi _r)} 
\right) - \frac{\partial {\cal L}}{\partial \varphi _r}\right] 
\delta \varphi _r
\, . 
\label{i} 
\end{equation}

Now, using the Euler-Lagrange equations

\begin{equation}
~^{4+D}\nabla _A\left(
\frac{\partial {\cal L}}{\partial (\partial _A\varphi _r)} 
\right) - \frac{\partial {\cal L}}{\partial \varphi _r} = 0 
\label{l} 
\end{equation}

and observing
that, for ${\cal M}^4\times \Sigma ^D$, we have
$~^{4+D}\nabla _A e^A_{(B)} = 0$
(here $~^{4+D}\nabla _A$ denotes the covariant derivative with
respect to the metric $J_{AB}$), equation (\ref{i})
rewrites as

\begin{equation}
~^{4+D}\nabla _A\left( 
\frac{\partial {\cal L}}{\partial (\partial _A\varphi _r)} 
\partial _{C} \varphi _r e^{C}_{(B)}
- {\cal L}e^{A}_{(B)}\right) = 0 
\, . 
\label{m} 
\end{equation}

This continuity equation leads to the conserved
4+D-dimensional fields momentum 

\begin{equation}
P_{(A)} = \int _{{\cal E}^3\times \Sigma ^D} d^3xd^D\theta
\left\{
\Pi _r\partial _{B} \varphi _r e^{B}_{(A)}
- {\cal L}e^0_{(A)} \right\}
\, , 
\label{n} 
\end{equation}

where $\Pi _r$ corresponds to the conjugate momenta of
$\varphi _r$ and behaves as 4+D-dimensional
densities
of weigh $1/2$.\\ 
The dependence of the fields $\varphi _r$ on the
extra-coordinates $\theta ^m$ has to be
in the form of a ``phase'' factor'', because
their matrix elements must not depend
on them, i.e. we assume the structure 

\begin{equation}
\varphi _r = \frac{1}{\sqrt{V}}
e^{-i\tau _{rs}(\theta ^m)}
\phi _s(x^{\mu }) \;
\Rightarrow 
\Pi _r = \frac{1}{\sqrt{V}}
\pi _s(x^{\mu })
e^{i\tau _{sr}(\theta ^m)} \;
\, ,
\label{p} 
\end{equation}

being $\phi _r$ and $\pi _r$ 4-dimensional conjugate variables,
while $\tau _{rs}(\theta ^m)$ denoting generic
Hermitian matrices
compatible with the symmetries of $\Sigma ^D$.\\ 
The substitution of these expressions into equation (\ref{n}) 
provides the outcomings

\begin{eqnarray}
\label{q} 
Q_{\mu } \equiv P_{\mu } = \int _{{\cal E}^3} d^3x
\left\{
\pi _r\partial _{\mu } \phi _r -
{\cal L}\delta ^0_{\mu } \right\} \\ 
Q_{(m)} = -\frac{i}{V}\int _{{\cal E}^3\times \Sigma ^D}
\sqrt{{\cal K}}
d^3x d^D\theta 
\left\{ \pi _r
\xi ^n_{(m)}\partial _n\tau _{rs}(\theta ^m)\phi _s \right\}
\, ,
\end{eqnarray}

above ${\cal K}$ refers to the determinant of the
extra-dimensional metric.\\
We see that
the 4-dimensional component of the conserved current
corresponds to the ordinary 4-momentum vector.
Furthermore we take the position
$\tau _{rs}(\theta ^m) =
T_{(p)\; rs}\lambda ^{(p)}_{(q)}\Theta ^{(q)} (\theta ^m)$, 
being $T_{(p)\; rs}$ the gauge generators, $\lambda ^{(p)}_{(q)}$ a
constant matrix to be determined and
$\Theta ^{(q)}$ expandible
in the harmonic functions of $\Sigma ^D$ \cite{SS82};
hence the charges
$Q_{(m)}$ rewrite

\begin{equation}
Q_{(m)} = -i\sqrt{\frac{c^3}{16\pi ~^4G}}
\int _{{\cal E}^3}
d^3x  \left\{ \pi _r
T_{(m)\; rs}\phi _s \right\}
\, .
\label{r} 
\end{equation}

as soon as we identify the matrix ${{\lambda }^{-1}}^{(p)}_{(q)}$
with the quantities 

\begin{equation}
{{\lambda }^{-1}}^{(p)}_{(q)} = 
\frac{\sqrt{16\pi ~^4G}}{c^{3/2}V}\int _{\Sigma ^D}
\sqrt{{\cal K}}
d^D\theta \left\{
\xi ^m_{(q)}\partial _m\Theta ^{(q)}\right\}
\, , 
\label{rr} 
\end{equation}

where by 
${{\lambda }^{-1}}^{(p)}_{(q)}$
we denote the inverse matrix of $\lambda ^{(p)}_{(q)}$ 
(i.e. ${{\lambda }^{-1}}^{(p)}_{(r)}
\lambda ^{(r)}_{(q)} = \delta ^{(p)}_{(q)}$).
Equation (\ref{r}) coincides
(apart from a factor $-\gamma$ which do not affect
the conservation law)
with the conserved quantities (\ref{b})
and therefore it shows how, in a
Kaluza-Klein theory, the charges associated to an Abelian or
non-Abelian gauge theory come out from the extra-components
of the fields momentum vector.

However, 
under the same assumption, equations (\ref{i}) and (\ref{p})
provide, for the infinitesimal coordinates transformation 
$x^{\mu ^{\prime }} = x^{\mu },\; 
\theta ^{m ^{\prime }} = \theta ^m +
\delta \omega ^{(p)}\xi ^m_{(p)}$, the following gauge
transformation on $\phi_r$

\begin{equation}
\phi ^{\prime }_r = \left(\delta _{rs}
- i\delta \omega ^{(q)}\lambda _{(q)}^{(n)}\xi ^m_{(n)}
\partial _m\Theta ^{(p)}T_{(p)\; rs}\right) \phi _s
\, .
\label{s} 
\end{equation}

Expression (\ref{s}) becomes equivalent
to (\ref{a}) only if we require

\begin{equation}
\xi ^m_{(n)}\partial _m \Theta ^{(p)} = \delta ^{(p)}_{(n)}
\Rightarrow
\xi ^{(n)}_m = \partial _m \Theta ^{(n)}
\, . 
\label{t} 
\end{equation}

This result is equivalent to the vanishing of all the
structures constants $C^{(p)}_{(q)(r)}$. Thus the
extra-dimensional components of the 4+D-momentum vector
become those 4-dimensional ones of a 
conserved charge in correspondence to Abelian or non-Abelian
gauge theories; but
the gauge transformation of the fields is induced by the
translation along the extra-dimensions only for
the Abelian case, 
when the line element of $\Sigma ^D$ can be reduced to the
Euclidean
one, i.e. $~^{D}dl^2=\sum _{m=1}^{D}(d\Theta ^m)^2$. 

\section{DIRAC ALGEBRA}

In a flat 4+D-dimensional Minkowski space ${\cal M}^{4+D}$,
the Lagrangian density of a set of
massless spinor fields
(the presence of a mass term does not affect the below
analysis, while the chirality of the spinors is not
addressed here \cite{W83})
$\Psi _r(X^A)$ takes the form

\begin{equation}
{\cal L}_{\Psi } = \frac{i}{2}\left(
\partial _A\bar{\Psi _r}\gamma ^A\Psi _r -
\bar{\Psi }\gamma ^A\partial _A\Psi _r\right)
\, , 
\label{u} 
\end{equation}

where by $\gamma ^A$ ($\bar{\Psi }=\Psi ^+\gamma ^0$) we
denote the Dirac matrices, satisfying the anti-commutation
relations

\begin{equation}
\left\{ \gamma _A\; ,\; \gamma _B\right\} = 2I\eta _{AB}
\, , 
\label{v} 
\end{equation}

being $I$ the identity matrix and $\eta _{AB}$ the
MinkoWskian metric.\\ 
On a curved 4+D-dimensional spacetime, the Dirac matrices
become functions on the manifold and have to be taken in
the form $\gamma _A(X^B) = \gamma _{(B)}e^{(B)}_A$, being the
4+D-bein components equal to the constant matrices (\ref{u}).\\
Thus, on a curved spacetime,
the relation (\ref{v}) rewrites as

\begin{equation}
\left\{ \gamma _A(X^C)\; ,\; \gamma _B(X^C)\right\} =
2Ij_{AB}(X^C)
\, , 
\label{vxv} 
\end{equation}

In correspondence to the vectors (\ref{f}) and (\ref{g}) the
matrices $\gamma ^{\mu } = \gamma ^{(\nu )}u^{\mu }_{(\nu )}$
and $\gamma _{\mu }=\gamma _{(\nu )}u^{(\nu )}_{\mu }$
define the appropriate 4-dimensional Dirac algebra with
respect to the 4-metric
$g_{\mu \nu } \equiv \eta _{(\rho )(\sigma )}
u^{(\rho )}_{\mu }u^{(\sigma )}_{\nu }$.

On a curved spacetime $~^{4+D}{\cal V}$,
the Lagrangian density (\ref{u})
rewrites as 

\begin{equation}
{\cal L}_{\Psi }^{Curv} = \frac{i}{2}\left(
{\cal D} _A\bar{\Psi _r}\gamma ^A\Psi _r -
\bar{\Psi }\gamma ^A{\cal D}_A\Psi _r\right)
\, \quad
{\cal D}_A \equiv \partial _A \pm \Gamma _A
\, , 
\label{x} 
\end{equation}

where (-) and (+) refer respectively to the application
of the spinor derivative $D_A$ on $\Psi $ and $\bar{\Psi }$.
The quantity $\Gamma _A$ is a kind of
``gauge connection'' for the
Lorentz group and reads

\begin{eqnarray}
\label{y} 
\Gamma _A = \Sigma ^{(B)(C)}\Omega _{(B)(C)\; A}\\
\Omega _{(B)(C)\; A} \equiv e^D_{(C)} ~^{4+D}\nabla _A
e_{(B)D}\\
\Sigma ^{(A)(B)}
\equiv \frac{1}{4} \left[\gamma ^{(A)}\; , \; \gamma ^{(B)}\right] 
\, . 
\end{eqnarray}

Above $\Sigma ^{(A)(B)}$ is the generator of the
Lorentz group in the spinor representation, while 
$\Omega _{(B)(C)\; A} $ play the role of the corresponding six
gauge vectors (which in the Einstein theory can be expressed via
the bein vectors $e^{(B)}_A$).\\ 
In agreement to the Spontaneous Compactification idea,
within a Kaluza-Klein theory, the 4+D-dimensional Lorentz
group is broken (near the ``vacuum state``
${\cal M}^4\times \Sigma ^D$)
into the 4-dimensional Lorentz one plus the D-dimensional
translation group. Therefore, in this              
framework, the
bein component of the quantity (\ref{y})
(i.e. $\Gamma _({A)} \equiv \Gamma _Be^B_{(A)}$) has to
admit only the 4-dimensional term
of the form (\ref{y}), i.e. 
$\Gamma _{(\mu )} = \Sigma ^{(\nu )(\rho )}
u^{\alpha }_{(\rho )}u^{\beta }_{(\mu )}
~^4\nabla _{\beta }u_{\alpha (\nu )}$.\\
The form of the bein component $\Gamma _{(m)}$ will be
determined in the next Section by requiring that, on
${\cal V}^4\times \Sigma^D$, the Lagrangian density (\ref{x})
provides the gauge coupling between the 4-spinors
and the Yang-Mills fields.

\section{GEOMETRIZATION OF THE GAUGE CONNECTION}

We start from the following action for a set of
4+D-dimensional spinor fields

\begin{equation}
S_{\Psi }^{Curv} = \frac{i}{2c}
\int_{{\cal V}^4\times \Sigma ^D}d^4xd^D\theta 
\left\{ E\left(
{\cal D} _{(A)}\bar{\Psi _r}\gamma ^{(A)}\Psi _r -
\bar{\Psi }\gamma ^{(A)}{\cal D} _{(A)}\Psi _r\right) \right\} 
\, , 
\label{z} 
\end{equation}

being $E\equiv det e^{(B)}_A$. Recalling that
$\gamma ^{(A)}D_{(A)} \equiv \gamma ^{(A)}\left(
\partial _{(A)} \pm \Gamma _{(A)}\right)$, we can split the
above action via the framework of Sections $4$ and  $5$.\\
In fact we have

\begin{equation}
\partial _{(A)} \equiv e^B_{(A)}\partial _B
\,,
\label{a1x} 
\end{equation}

 and hence, by (\ref{g}), the following relations outcome

\begin{eqnarray}
\label{a2x} 
\partial _{(\rho )} = u^{\mu }_{(\rho )}\partial _{\mu } -
\gamma u_{(\rho )}^{\mu }A^{(p)}_{\mu }\xi _{(p)}^m\partial _m \equiv 
~^4\partial _{(\rho )} -
\gamma u_{(\rho )}^{\mu }A^{(p)}_{\mu }\xi _{(p)}^m\partial _m \\
\partial _{(m)} = \xi _{(m)}^p\partial _p \equiv ~^D\partial _{(m}) 
\, ;
\end{eqnarray}

above, by $~^4\partial _{(a)}$ and
$~^D\partial _{(m)}$, we denoted the directional
derivatives respectively on ${\cal V}^4$ and $\Sigma ^D$.\\
Furthermore we get

\begin{eqnarray}
\label{a3x} 
\gamma ^{(A)}\Gamma _{(A)} =
\gamma ^{(\mu )}\Sigma ^{(\rho )(\sigma )}
\Omega _{(\rho )(\sigma )(\mu )} +
\gamma ^{(m)}\Gamma _{(m)}\\
\partial _{(m)}\Psi _r = -i\xi^n_{(m)}T_{(p)\; rs}
\lambda ^{(p)}_{(q)}\partial _n\Theta ^{(q)}\Psi _s\\ 
\partial _{(m)}\bar{\Psi }_r =
i\xi^n_{(m)}
\bar{\Psi }_sT_{(p)\; sr}
\lambda ^{(p)}_{(q)}\partial _n\Theta ^{(q)} \\
\, ,
\end{eqnarray}

as well as we have
$\bar{\Psi _r}\Psi _r = \bar{\psi _r}\psi _r/V$,
being, in agreement to (\ref{p}),
$\{ \bar{\psi }(x^{\mu })\, , \psi (x^{\mu })\}$ the
4-dimensional spinor fields. 

Putting together these results and 
taking in (\ref{z}), the integral over the
extra-coordinates, it provides the following
4-dimensional action

\begin{eqnarray}
\label{a4x} 
S_{\Psi }^{Curv} = \frac{1}{c}
\int_{{\cal V}^4}d^4x 
U\left\{ \frac{i}{2}\left(
~^4{\cal D} _{(\mu )}\bar{\psi _r}\gamma ^{(\mu )}\psi _r -
\bar{\psi }\gamma ^{(\mu )}~^4{\cal D} _{(\mu )}\psi _r\right)
+ {\cal L}_{int}\right\} \\
{\cal L}_{int} = 
g\bar{\psi _r}\gamma^{(\nu )}
u_{(\nu )}^{\mu }
A^{(m)}_{\mu }
T_{(m)\; rs}\psi _s
- \bar{\psi }_r\gamma ^{(m)}\left(
\sqrt{\frac{c^3}{16\pi ~^4G}}  
T_{(m)\; rs}
- iV\Gamma _{(m)}\right) \psi _s 
\, , 
\end{eqnarray}

where $U\equiv detu^{(a)}_{\mu }$ and $~^4{\cal D}_{(\mu )}$ denotes the
ordinary 4-dimensional spinor derivative projected on the
4-bein.\\
In the above action, the first two terms
provide the free 4-spinor fields, while those ones
in $S_{int}$ correspond respectively to the desired
gauge coupling
and to a contribution which is not experimentally
detected; to remove such a term we need the choice
$\Gamma _{(m)} =
-i\frac{\sqrt{c^3/16\pi ~^4G}}{V}T_{(m)\; rs}I$.

At last, the above action rewrites 

\begin{eqnarray}
S_{\Psi }^{Curv} = \frac{1}{c}
\int_{{\cal V}^4}d^4x 
U\left\{ \frac{i}{2}\left(
\left( ~^4{\cal D} _{(\mu )}
- igT_{(m)\; rs}A^{(m)}_{(\mu )}\right)
\bar{\psi _r}\gamma ^{(\mu )}\psi _s -
\bar{\psi }_r\gamma ^{(\mu )}
\left( ~^4{\cal D} _{(\mu )}
+ igT_{(m)\; rs}A^{(m)}_{(\mu )}\right)
\psi _s\right)
\right\}
\label{a4x4}
\, .
\end{eqnarray}

We see how, after the dimensional reduction,
the 4-dimensional action contains the correct gauge
coupling, which appears as a consequence of the
geometrical nature of the Yang-Mills fields within
a Kaluza-Klein theory.

\section{BRIEF CONCLUDING REMARKS}

Putting together the geometrization of the
Yang-Mills fields, performed in the usual
 Kaluza-Klein approach, with the result here
 obtained, we see that, starting from the
4+D-dimensional gravity-matter action
$~^{4+D}S = ~^{4+D}S_{E-H} + 
~^{4+D}S_{\Psi }^{curv}$, after the dimensional
reduction, we get a 4-dimensional action describing
all the appropriate bosonic and fermionic components
with their relative couplings, i.e. (with obvious notation) 
$~^{4}S = ~^{4}S_{E-H, \Lambda } + ~^{4}S_{Y-M}^{curv} + 
~^{4}S_{\psi }^{curv} + ~^{4}S_{int}$.

The above analysis shows how, in the framework
of a Kaluza-Klein theory, not only the Yang-Mills
fields can be geometrized, but also their gauge
couplings outcome from the splitting of
the geometrical terms contained in the matter action.\\
The assumptions at the ground of 
our point of view are supported by the interpretation
of the extra-dimensional components of the fields
momentum in terms of gauge charges.
However a difference has to be emphasized between
the Abelian and non-Abelian case; in fact,
while for an abelian group the translations along the
extra-dimensions induce directly the 4-dimensional
gauge transformations, the latter one,
in the non-Abelian case can be recognized only
in the structure of the (dimensionally reduced)
4-dimensional action.
In this sense, the geometrization of the non-abelian
gauge connection privilages the Lagrangian
representation of the dimensional reduction with
respect to an approach based on the field equations.
Indeed, if we write down the 4+D-dimensional Dirac
equation and split it in terms of 4-dimensional variables,
then to get the right gauge coupling terms it
would be necessary to carry out an additional
integration 
over the extra-dimensional coordinates; however,
this same picture would appear when splitting the
4+D-dimensional Einstein equations
toward the 4-dimensional
Einstein-Yang-Mills theory. Thus the different
behavior, above outlined, between the Abelian and non-Abelian
theories with respect to a geometrical interpretation,
is a general feature of the Kaluza-Klein approach
and it is not due to the specific assumptions
we addressed here.

\end{document}